
\input harvmac
\input epsf
\def \pa{\partial}
\def\npb{{Nucl.\ Phys.\ }{\bf B}}
\def\physrep{Phys.\ Reports\ }
\def\plb{{Phys.\ Lett.\ }{\bf B}}
\def\prd{{Phys.\ Rev.\ }{\bf D}}
\def\prl{Phys.\ Rev.\ Lett.\ }
\def\ptp{Prog.\ Th.\ Phys.\ }

\def\half{{\textstyle{1\over2}}} 
\def\frak#1#2{{\textstyle{{#1}\over{#2}}}}
\def\frakk#1#2{{{#1}\over{#2}}}
\def\taub{{\overline{\tau}}}
\def\bbar{{\overline{b}}}
\def\tbar{{\overline{t}}}
\def\sy{supersymmetry}
\def\sic{supersymmetric}

\def\lf{16\pi^2}

\def\GeV{{\rm GeV}}
\def\TeV{{\rm TeV}}
\def \in{\leftskip = 40 pt\rightskip = 40pt}
\def \out{\leftskip = 0 pt\rightskip = 0pt}
{\nopagenumbers
\line{\hfil LTH 450}
\line{\hfil hep-ph/9903365}
\line{\hfil Revised Version}
\vskip .5in    
\centerline{\titlefont Non-standard soft supersymmetry breaking
}
\vskip 1in
\centerline{\bf I.~Jack and D.R.T.~Jones}
\medskip
\centerline{\it Dept. of Mathematical Sciences,
University of Liverpool, Liverpool L69 3BX, UK}
\vskip .3in

We explore a more general class of  soft supersymmetry-breaking  masses
and interactions than that usually considered, both  in general and in
the MSSM context, where our  results for the 
one-loop $\beta$-functions correct some errors in the literature. 
We identify a new class of one-loop finite \sic\ theories.

\Date{March 1999}

\newsec{The new soft breakings}
The minimal \sic\ standard model (MSSM) consists of a supersymmetric 
extension of the standard model, with the addition of a number  of
dimension 2 and dimension 3 \sy-breaking mass and interaction terms.  
It became popular when it was  demonstrated that such a structure is a
natural  consequence of supergravity when \sy\ is broken in a hidden
sector. (For a  review, see Ref.~\ref\nilles{ H.-P.~Nilles, \physrep 
C110 (1984) 1}.) The purpose of this paper  is a preliminary exploration
of the consequences of a more general  set of \sy-breaking terms. 
For a general ${\cal N} =1$ theory, let us  write
\eqn\Ac{
L=L_{\rm SUSY}+L_{\rm SOFT}.}
Here $L_{\rm SUSY}$ is the Lagrangian for  the supersymmetric
gauge theory, containing the gauge multiplet $\{A_{\mu},\lambda\}$ ($\lambda$ 
being the gaugino) and a matter multiplet 
$\{\phi_i,\psi_i\}$ transforming as a  
representation $R$ of the gauge group $\cal G$.  
We assume a superpotential of the form   
\eqn\Aae{
W=\frak{1}{6}Y^{ijk}\phi_i\phi_j\phi_k.}

A renormalisable superpotential will in general also contain quadratic
and  linear terms. We suppose that there are no gauge singlet fields so
there  is no linear term; and as will become clear below, we do not need
  an explicit quadratic term  because such a term will be included as a 
special case from our new soft breakings.

The soft terms usually considered are those contained 
in the following Lagrangian: 
\eqn\Aafo{
L^{(1)}_{\rm SOFT}=(m^2)^j{}_i\phi^{i}\phi_j+
\left(\frak{1}{6}h^{ijk}\phi_i\phi_j\phi_k+\half b^{ij}\phi_i\phi_j
+ \half M\lambda\lambda+{\rm h.c.}\right).}
Indeed, in the MSSM context one often sees the (incorrect) assertion that 
$L^{(1)}_{\rm SOFT}$ contains all possible soft terms
\foot{For a recent honourable exception and a nice MSSM review, see 
Ref.~\ref\martin{S.~Martin, hep-ph/9709356}}. 
The designation ``soft'' refers to the fact that the inclusion of 
$L^{(1)}_{\rm SOFT}$ breaks \sy\ but does not introduce quadratic
divergences\ref\girgris{D.M.~Capper, J. Phys. {\bf G}3 (1977) 731\semi
L.~Girardello and M.T.~Grisaru, \npb 194 (1982) 65}, 
and  is hence said to preserve 
naturalness\foot{In a $U_1$ theory  naturalness also requires 
$\tr Y = 0$, where $Y$ is the $U_1$
hypercharge\ref\fnrs{W.~Fischler et al, \prl 47 (1981) 757}  
}. However in the case of a wide range of theories 
there are further possible dimension 3 terms which preserve
naturalness, as follows:
\eqn\Aaf{
L^{(2)}_{\rm SOFT}=\half r_{i}^{jk}\phi^i\phi_j\phi_k
+\half m_F{}^{ij}\psi_i\psi_j
+ m_A{}^{ia}\psi_i\lambda_a
+{\rm h.c.}}
The $m_A$ term (first discussed in Ref.~\ref\jmy{D.R.T. Jones, L.
Mezincescu and Y.-P. Yao, \plb148 (1984) 317})  is only possible 
given adjoint  matter fields; not a feature of the MSSM, but often 
encountered in GUTs. The reason the terms exhibited in Eq.~\Aaf\ do not
appear   in the  classification of Ref.~\girgris\  is that {\it in
general\/} they  engender quadratic divergences. These divergences are
in scalar tadpoles,  and hence absent if there are no gauge
singlet matter fields;  as is the case in the MSSM\foot{Although
singlets are a popular addition  in not-so-minimal models; recently in
the context of (Dirac) neutrino masses}. Thus a truly
model-independent  approach to the MSSM should include terms of the form
shown in Eq.~\Aaf.

\newsec{The one-loop $\beta$-functions}

We now present the one-loop $\beta$-functions for 
$L_{\rm SOFT} = L^{(1)}_{\rm SOFT} + L^{(2)}_{\rm SOFT}$. 
The $\beta$-functions  for scalar masses and interactions may be
calculated  using the following equation for the  tree scalar potential
$V_0$:   
\eqn\vrg{ \left[\sum_I\beta_I\frakk{\pa}{\pa\lambda_I}-
(\phi_i\gamma_L{}^i{}_j \frakk{\pa}{\pa\phi^j}  + {\rm h.c. })\right]V_0
= \frakk{1}{32\pi^2} \rm{STr}\, {\cal M}^4.} 
The sum  over $I$ includes all
masses and couplings, and ${\rm STr}$ stands for the  usual
spin-weighted trace.   (Note that $\gamma_L$ is the Landau gauge scalar
anomalous  dimension, which differs from the chiral superfield anomalous
dimension, $\gamma$.)   This equation, in fact, was  employed in
Ref.~\jmy\ to seek and classify one-loop finite  theories.  In the case
of  the $\beta$-functions for the  fermion mass terms the explicit
calculation is very simple.  

The one-loop results for the gauge coupling $\beta$-function $\beta_g$
and  for $\gamma$ are:

\eqn\Aab{
\lf\beta_g =g^3Q \quad\hbox{and}\quad 
\lf\gamma^{i}{}_j=P^i{}_j,}
where
\eqn\Aac{ 
Q=T(R)-3C(G),\quad\hbox{and}\quad
P^i{}_j=\half Y^{ikl}Y_{jkl}-2g^2C(R)^i{}_j.}
Here
\eqn\Aaca{
T(R)\delta_{ab} = \Tr(R_a R_b),
\quad C(G)\delta_{ab} = f_{acd}f_{bcd} \quad\hbox{and}\quad 
C(R)^i{}_j = (R_a R_a)^i{}_j,}
and as usual $Y_{ijk}^* = Y^{ijk}$ etc. 
For the new soft terms from Eq.~\Aaf\ we find:
\eqna\mfij$$\eqalignno{
\lf(\beta_{m_{F}})_{ij} &= P^k{}_i m_{Fkj} + P^k{}_j m_{Fik},  &\mfij a\cr
\lf(\beta_{m_{A}})_{ia} &= P^j{}_i m_{Aja} + g^2Q m_{Aia}, &\mfij b\cr}$$
and
\eqn\Ac{\eqalign{
\lf(\beta_r)^{jk}_i &=  \half P^l{}_i r^{jk}_l + P^k{}_l r^{jl}_i 
+\half r^{mn}_i Y_{lmn}Y^{ljk}+2r^{mj}_l Y_{imn}Y^{kln} 
+ 2g^2 r^{jk}_l C(R)^l{}_i\cr &+2g^2r^{mj}_l (R_a)^k{}_i (R_a)^l{}_m
-2m_{Flm}Y^{mnj}Y^{plk}Y_{npi} -4g^2m_{Fil}C(R)^l{}_mY^{mjk}\cr 
&- 4g\sqrt{2} \left[g^2C(G) m_A^{ja}(R_a)^k{}_i
+ (R_a)^j{}_l Y^{lmk}Y_{mni}m_A^{na}\right] \quad +(k\leftrightarrow j).\cr}}
For the original soft terms in Eq.~\Aafo\ we find
\eqna\Acc$$\eqalignno{
\lf\beta_h^{ijk}&=U^{ijk}+U^{kij}+U^{jki}, &\Acc a\cr
\lf\beta_b^{ij}&=V^{ij}+V^{ji}, &\Acc b\cr
\lf[\beta_{m^2}]^i{}_j&=W^i{}_j, &\Acc c\cr
\lf\beta_M&=2g^2QM, &\Acc d\cr
}$$
where
\eqna\Aoldm$$\eqalignno{
U^{ijk}&=h^{ijl}P^k{}_l+Y^{ijl}X^k{}_l, &\Aoldm a\cr
V^{ij}&=b^{il}P^j{}_l+ r^i_{lm}h^{jlm}
+r^{im}_l r^{jl}_m -m_{Fkl}Y^{ilm}m_{Fmn}Y^{jnk} \cr &
+4g^2Mm_F^{ik}C(R)^j{}_k -4g^2C(G)m_A^{ia}m_A^{ja},
&\Aoldm b\cr 
W^i{}_j&= \half Y_{jpq}Y^{pqn}(m^2)^i{}_n
+\half Y^{ipq}Y_{pqn}(m^2)^n{}_j
+2Y^{ipq}Y_{jpr}(m^2)^r{}_q
+h_{jpq}h^{ipq}\cr & 
+r^{kl}_{j}r_{kl}^i +2r^k_{jl}r^{il}_k 
-4(m_F^{kl}m_{Flm}+ m_{Ama}m_A^{ka})Y^{imn}Y_{jkn}\cr
&-8g^2(MM^*C(R)^i{}_j + m_F^{kl}m_{Fjk}C(R)^i{}_l +C(G)m_A^{ia}m_{Aja}
+(R_a R_b)^i{}_j m_{Aka}m_A^{kb})\cr
&-4\sqrt{2}g(Y^{iml}m_{Fmn}(R_a)^n{}_j m_{Ala}
+Y_{jml}m_F^{mn}(R_a)^i{}_n m_A^{la})&\Aoldm c \cr}$$
with
\eqn\Aab{
X^i{}_j=h^{ikl}Y_{jkl}+4g^2MC(R)^i{}_j.}

In the expression corresponding to Eq.~\Acc{c}\ in 
Ref~\ref\jj{I.~Jack and
D.R.T.~Jones, \plb 333 (1994) 372}, 
there is an additional contribution of the form $g^2(R_a)^i{}_j\Tr[R_am^2]$. 
This term arises only for $U(1)$ and amounts to a renormalisation of 
the linear $D$-term that is allowed in that case. 

In the special case when $m_A^{ia} = h^{ijk} = M = b^{ij} = 0$, 
$m_F = \mu$, $r_i^{jk} = Y^{jkl}\mu_{il}$ and 
$(m^2)^i{}_j = \mu^{il}\mu_{jl}$ then the theory becomes 
supersymmetric, with 
\eqn\muij{\lf(\beta_{\mu})_{ij} = P^k{}_i \mu_{kj} + P^k{}_j \mu_{ik}.}
It is easy to check that Eqs.~\mfij{a}, \Ac, \Aoldm{c}\ 
are consistent with this result. 
In the case $m_A^{ia}=0$, $m_F = \mu$, $r_i^{jk} = Y^{jkl}\mu_{il}$  
and $(m^2)^i{}_j \to (m^2)^i{}_j + \mu^{il}\mu_{jl}$  
our results reduce to the usual soft $\beta$-functions, as
given in Ref.~\jj\ (see also 
Ref.~\ref\mv{S.P.~Martin
and M.T.~Vaughn, \prd50 (1994) 2282\semi
Y.~Yamada, \prd50 (1994) 3537\semi
I.~Jack et al, \prd50 (1994) R5481}). 
It is easy to see that 
this corresponds to the inclusion of a term 
$\half \mu^{ij}\phi_i\phi_j$ in the superpotential. 
This is why 
we do not need  to include such a term  in Eq.\Aae. 
Indeed, a plausible common origin for the new and 
usual soft terms would form the basis for 
a solution to the so-called 
``$\mu$ problem''.  

An interesting special case is provided by one-loop finite theories
such that $P = Q = 0$. Theories with $r_i^{jk} = m_F = m_A = 0$ were 
considered in Ref.~\jmy; but there are other possibilities.  
Note that we have immediately that 
$\beta_{m_{F}} = \beta_{m_{A}} = 0$ and if we 
set\foot{One loop finite 
theories with ${\cal N}=2$ \sy\ and non\-zero $r^{jk}_i$ 
were constructed in Ref.~\ref\frluced{
J.M.~Fr\`ere, L.~Mezincescu and Y-P. Yao, 
\prd 29 (1984) 1196; \prd 30 (1984) 2238}} 
\eqn\rfin{r_i^{jk}  = \sqrt{2}g\left[(R_a)^j{}_i m_A^{ka} 
+ (R_a)^k{}_i m_A^{ja}\right]}
and $m_F = 0$, we find that $\beta_r = \beta_b = 0$. 
If we additionally set
\eqn\mmfin{ m^{Aia} m_{Aja} = \rho\delta^i{}_j,  h = -MY, 
(m^2)^i{}_j = (2\rho + \frak{1}{3}MM^*)\delta^i{}_j
\quad\hbox{and}\quad 
C(R)^i{}_j = C(G) \delta^i{}_j,}
then we have 
$W^i{}_j = X^i{}_j = 0$ and one-loop finiteness. A theory 
that can satisfy these constraints is one with ${\cal G} = SU(N)$,  
three adjoint matter superfields and the 
superpotential\ref\timluca{D.R.T. Jones and L.~Mezincescu, \plb138 (1984) 293} 
\eqn\nfourd{
W = gN\sqrt{\frakk{2}{N^2 - 4}}d^{abc}\phi^a_1\phi^b_2\phi^c_3,}
where the unbroken theory has the field content of 
${\cal N}=4$, but no higher \sy.   
 
\newsec{The MSSM} We now turn to the case of the MSSM, in the 
approximation where we retain only the third generation  Yukawa
couplings. In this context, in fact, the existence of both   $r^{jk}_i$
and $m_F$-type terms was entertained in a pioneering  paper on
the  MSSM\ref\inetal{K.~Inoue et al, \ptp 67 (1982) 1889;  erratum {\it
ibid\/} 70 (1983) 330}\ so we adopt some of their notation for
convenience of comparison.  Thus we write  
\eqn\wmssm{W = \lambda_t H_2
Q \tbar + \lambda_b H_1 Q \bbar  + \lambda_{\tau}H_1 L \taub,}

\eqn\smssmb{\eqalign{ L^{(1)}_{\rm SOFT} &= \sum_{\phi}
m_{\phi}^2\phi^*\phi + \left[m_3^2 H_1
H_2 + \sum_{i=1}^3\half M_i\lambda_i\lambda_i  + {\rm h.c. }\right]\cr 
&+ \left[m_{10}\lambda_t H_2 Q \tbar  +
m_8\lambda_b H_1 Q \bbar  + m_6\lambda_{\tau}H_1 L \taub 
+ {\rm h.c. }\right]\cr}}

and 
\eqn\smssmb{ L^{(2)}_{\rm SOFT} = m_4 \psi_{H_1}\psi_{H_2} + 
m_9\lambda_t H_1^* Q \tbar 
+ m_7\lambda_b H_2^* Q \bbar 
+ m_5\lambda_{\tau}H_2^* L \taub   + {\rm h.c. }}
Nowadays $m_{6,8,10}$ are usually 
written $A_{\tau,b,t}$ respectively. 
We note en passant that if R-parity violation is allowed then, as is 
well known, there are various additional terms allowed in $W$; the 
extra allowed terms of the $\phi^2\phi^*$ and $\psi\psi$-type are 
as follows (for one generation):
\eqn\smssmb{ L^{(2)RPV}_{\rm SOFT}
=  \rho_1 L^* Q \tbar + \rho_2 H_2^* H_1\taub + 
m_{\rho} \psi_{L}\psi_{H_2} + {\rm h.c. },}
but we do not pursue this possibility here. 

It is straightforward to show from our results that
\eqna\fourb$$\eqalignno{
\lf\beta_{m_1^2} &= 
2\lambda_{\tau}^2 (m_1^2 + m_6^2 + m_L^2 + m_{\taub}^2)
+6\lambda_b^2 (m_1^2 + m_8^2 + m_Q^2 + m_{\bbar}^2)\cr
&+6\lambda_t^2 m_9^2 
-8C_Hm_4^2 -6g_2^2M_2^2-2{g'}^2M_1^2, &\fourb a\cr
\lf\beta_{m_2^2} &= 
6\lambda_t^2 (m_2^2 + m_{10}^2 + m_Q^2 + m_{\tbar}^2)
+2\lambda_{\tau}^2 m_5^2+6\lambda_b^2 m_7^2\cr
&-8C_Hm_4^2 -6g_2^2M_2^2-2{g'}^2M_1^2, &\fourb b\cr
\lf\beta_{m_3^2} &= 
(\lambda_{\tau}^2 + 3\lambda_b^2 + 3\lambda_t^2)m_3^2 
+ 2\lambda_{\tau}^2 m_5 m_6 + 6\lambda_b ^2 m_7 m_8 + 6\lambda_t^2 m_9 m_{10} 
\cr &-4 C_Hm_3^2 +6g_2^2m_4M_2 +2{g'}^2M_1m_4, &\fourb c\cr
\lf\beta_{m_4} &= (\lambda_{\tau}^2 + 3 \lambda_b^2 
+ 3\lambda_t^2- 4 C_H)m_4
, &\fourb d\cr
\lf\beta_{m_5} &= (\lambda_{\tau}^2 - 3 \lambda_b^2 + 3\lambda_t^2)m_5
+6m_7\lambda_b^2 +
(4m_5 - 8m_4)C_H, &\fourb e\cr
\lf\beta_{m_6} &= 8\lambda_{\tau}^2m_6 + 6\lambda_b^2 m_8 
+6g_2^2M_2 + 6{g'}^2M_1, &\fourb f\cr
\lf\beta_{m_7} &= (-\lambda_{\tau}^2 + 3 \lambda_b^2 + 5\lambda_t^2)m_7
+2m_5\lambda_{\tau}^2 +2\lambda_t^2 (m_9 -2m_4)\cr & + 
(4m_7 - 8m_4)C_H, &\fourb g\cr
\lf\beta_{m_8} &= 2\lambda_{\tau}^2m_6 + 12\lambda_b^2 m_8
+ 2\lambda_t^2 m_{10}\cr &+\frak{32}{3}g_3^2M_3 +6g_2^2M_2 
+ \frak{14}{9}{g'}^2M_1, &\fourb h\cr
\lf\beta_{m_9} &= (\lambda_{\tau}^2 + 5 \lambda_b^2 + 3\lambda_t^2)m_9
+2m_7\lambda_b^2 - 4m_4\lambda_b^2 +
(4m_9 - 8m_4)C_H, &\fourb i\cr
\lf\beta_{m_{10}} &= 2\lambda_b^2 m_8
+ 12\lambda_t^2 m_{10} +\frak{32}{3}g_3^2M_3 +6g_2^2M_2
+ \frak{26}{9}{g'}^2M_1, &\fourb j\cr
\lf\beta_{m_Q^2} &= 
2 X_b +2X_t -\frak{32}{3}g_3^2M_3^2  -6g_2^2M_2^2
- \frak{2}{9}{g'}^2M_1^2, &\fourb k\cr
\lf\beta_{m_{\tbar}^2} &= 
4X_t -\frak{32}{3}g_3^2M_3^2 
- \frak{32}{9}{g'}^2M_1^2, &\fourb l\cr
\lf\beta_{m_{\bbar}^2} &=
4X_b-\frak{32}{3}g_3^2M_3^2  
- \frak{8}{9}{g'}^2M_1^2, &\fourb m\cr
\lf\beta_{m_L^2} &=
2X_{\tau} -6g_2^2M_2^2
- 2{g'}^2M_1^2, &\fourb n\cr
\lf\beta_{m_{\taub}^2} &=
4X_{\tau}- 8{g'}^2M_1^2, &\fourb o\cr
\lf\beta_{M_i} &= 2b_iM_ig_i^2, &\fourb p\cr }$$
where $b_{1,2,3} = (33/5, -1, -3)$, ${g'}^2 = 3g_1^2/5$, 
$C_H = \frak{3}{4}g_2^2 + \frak{3}{20}g_1^2$ and 

\eqn\xdefs{\eqalign{
X_t &=   
\lambda_t^2 (m_Q^2 +  m_{\tbar}^2 + m_2^2 + m_9^2 + m_{10}^2  - 2m_4^2),\cr
X_b &=  \lambda_b^2 (m_Q^2 +  m_{\bbar}^2+  m_1^2 + m_7^2 + m_8^2 - 2m_4^2),\cr
X_{\tau} &=  
\lambda_{\tau}^2 (m_L^2 + m_{\taub}^2+  m_1^2 + m_5^2 + m_6^2 - 2m_4^2).\cr}}
The terms linear in the gaugino masses $M_i$ differ  by a
sign from Ref.~\inetal; this is a matter of convention.  The results for
$\beta_{m_7}$ and $\beta_{m_9}$, however, disagree. This 
appears to arise from the omission in Ref.~\inetal\ of 
some contributions which cancel in
the \sic\ limit. 

\newsec{IR fixed points}

In this section we discuss the RG evolution of $m_{4,5,7,9}$, with  emphasis
on possible fixed point (or quasi-fixed point) 
structure. In a recent paper\ref\jjir{I.~Jack and D.R.T.~Jones, 
\plb 443 (1998) 177},  we
showed that in a wide range of theories 
the existence of stable infra-red
fixed points for the Yukawa couplings implies 
stable infra-red fixed points for the $A$-parameters and soft scalar
masses.\foot{We first showed IR-focussing of soft
parameters for some GUTs in Ref.~\ref\fjj{P.~Ferreira,
I.~Jack and D.R.T.~Jones, \plb 357 (1995) 359}; see also 
Ref.~\ref\lross{ M.~Lanzagorta and G.G.~Ross, \plb 364 (1995) 63}. 
For recent analyses  in
the MSSM context, see  Ref.~\ref\aball{ S.~Abel and B.C.~Allanach, \plb
415 (1997) 371\semi G.H.~Yeghiyan, M.~Jurcisin and D.I.~Kazakov,
hep-ph/9807411 }\ (small $\tan\beta$) and  Ref.~\ref\kaz{M.~Jurcisin and
D.I.~Kazakov, Mod.~Phys.~Lett. A14 (1999) 671}\ (large $\tan\beta$).} We shall  see that
there is no such simple correspondence for the new soft interactions. 
 
It follows from Eq.~\fourb{}\ that  
there is a fixed point 
of the RG evolution 
such that\eqn\fxb{
\frakk{m_5}{m_4} = \frakk{m_7}{m_4} = \frakk{m_9}{m_4} = 1.}
This fixed point corresponds to the \sic\ limit for these parameters 
(supersymmetry is not fully restored since we 
do not have, for example, that $m_1/m_4 = 1$ is a fixed point).
An obvious question is whether Eq.~\fxb\ 
represents an {\it infra-red\/} fixed point of our theory, and if so 
whether fixed point (or, more likely, quasi-fixed-point) 
behaviour is exhibited in the standard evolution 
down to $M_Z$. The stability matrix for the 
evolution of $\frakk{m_5}{m_4}$, $\frakk{m_7}{m_4}$
and $\frakk{m_9}{m_4}$ is given by:
\eqn\matx{
S = \pmatrix{8C_H-6\lambda_b^2 & 6\lambda_b^2 & 0\cr
2\lambda_{\tau}^2 & 8C_H - 2\lambda_{\tau}^2 
+ 2\lambda_t^2 & 2\lambda_t^2 \cr
0 & 2\lambda_b^2 & 8C_H + 2\lambda_b^2\cr}}
which has eigenvalues
$8C_H, 8C_H + \Lambda_{1,2}$
where $\Lambda_{1,2}$ are the roots of the quadratic
\eqn\qudrtc{
\Lambda^2 - 2(\lambda_t^2 - \lambda_{\tau}^2 - 2\lambda_b^2)\Lambda
-4(3\lambda_b^2 + 3 \lambda_t^2 + \lambda_{\tau}^2)\lambda_b^2=0.}
Let us consider two special cases:

\subsec{The Quasi Fixed Point}

Suppose that 
we are near the quasi-infra-red fixed point (QIRFP) for $\lambda_t, 
\lambda_t(M_Z) \approx 1.1.$
This corresponds to $\tan\beta \approx 1.7$ and 
means  
we can neglect $\lambda_b$ and $\lambda_{\tau}$,
and it is easy to see that our fixed point is stable. 
With the Yukawa couplings and other soft parameters, one finds (given 
a stable fixed point) QIRFP behaviour rather than convergence 
to the fixed point. In this case, $m_7/m_4$ shows 
good fixed point convergence, while $m_9/m_4$ and $m_5/m_4$ approach much 
more slowly, with no marked QIRFP behaviour.  
If, for example, 
we have $m_5 = m_7 = m_9 = 0$ and $m_4 \neq 0$ at the gauge 
unification scale, $M_U$, then 
at $M_Z$ we find 
\eqn\fpres{
\frakk{m_5}{m_4} \approx  \frakk{m_9}{m_4} \approx 0.5, 
\quad \hbox{and}\quad \frakk{m_7}{m_4} \approx 0.9,}
whereas if we take $m_5 = m_7 = m_9 = 2m_4$ at $M_U$ then at $M_Z$ we find:
\eqn\trinb{  
\frakk{m_5}{m_4} \approx  \frakk{m_9}{m_4} \approx 1.5,
\quad \hbox{and}\quad \frakk{m_7}{m_4} \approx 1.1.}

The fact that $m_5$ and $m_9$ remain approximately equal 
is easy to understand from 
Eqs.~\fourb{e,i}\ using $\lambda_b\approx\lambda_{\tau}\approx 0$.


\subsec{Trinification}

There is a region of parameter space 
giving acceptable electro-weak breaking that corresponds to 
Yukawa trinification: 
$\lambda_t (M_U) \approx \lambda_b(M_U) \approx  
\lambda_{\tau}(M_U)\approx 0.6$ .
The corresponding value of $\tan\beta$ is $\tan\beta\approx 50$.
The two eigenvalues $8C_H + \Lambda_{1,2}$ are both positive at 
$M_U$ but one of them is negative at $M_Z$. Consequently 
we cannot anticipate that the fixed point (Eq.~\fxb) 
will be relevant. 
Indeed, taking    
 $m_5 = m_7 = m_9 = 0$ and $m_4 \neq 0$ at $M_U$, we find (at $M_Z$):
\eqn\trinb{
\frakk{m_9}{m_4} \approx  0.7,\quad  \frakk{m_7}{m_4} \approx 0.7, 
\quad \hbox{and}\quad \frakk{m_5}{m_4} \approx 0.4,}
whereas if we take $m_5 = m_7 = m_9 = 2m_4$ at $M_U$ then at $M_Z$ we find: 
\eqn\trinb{
\frakk{m_7}{m_4} \approx \frakk{m_9}{m_4} \approx 1.3,
\quad \hbox{and}\quad \frakk{m_5}{m_4} \approx 1.6,}
so in this case none of the parameters show fixed point behaviour, 
as expected. This time $m_7$ and $m_9$ remain approximately equal, and 
again this is easy to understand from Eqs~\fourb{g,i}, using 
$\lambda_b\approx\lambda_{\tau}\approx \lambda_t$.

We turn now to  a full running analysis of the theory, 
with the assumption that there is no explicit Higgs $\mu$-term.  

\newsec{RG evolution}

In general, if we admit these new soft breakings the  effect
is to enlarge the (already gargantuan) parameter space  of the MSSM.
This parameter space is customarily controlled  in the MSSM by
assumptions of unification for the soft scalar masses (to $m_0$), gaugino
masses   (to $M$) and $A$-parameters (to $A$). 
A distinctive possibility  within our
scenario is as follows: suppose we adopt this unification, 
the non-standard soft  terms are
present,  $m_{5,7,9}$ unify to $m_r$, 
{\it and there is no $\mu$-term in the superpotential}. 
In the  special
case that the soft terms satisfy $m_{4,5,7,9} = 0$,  this
corresponds to the MSSM without a $\mu$-term. 
Now in the standard running analysis, the Higgs potential minimisation 
is used to determine $m_3^2$ and $\mu^2$ (at $M_Z$). 
We are, however, constrained by the absence of a $\mu$ term and the fact that 
we are still requiring $m_1^2$ and $m_2^2$ to unify at $M_U$.

As discussed recently by 
Falk\ref\falk{T.~Falk, hep-ph/9902352}, the MSSM with a $\mu$ term such 
that $|\mu| < 0.4M$, say,  
is restricted to a very small region of parameter space at 
$m_0 >> M$. As a consequence, it is difficult to arrange for 
a Higgsino-like lightest neutralino. In our scenario, however, 
it turns out that the fact that $m_4$ and $m_r$ are 
``divorced'' from $\mu$ means we are able to achieve acceptable 
vacua with $m_4 \leq M$ while retaining unification for both 
scalar and gaugino masses. Values for $m_0$ are  lower than in the 
MSSM ($\mu = 0$) case but for an acceptable vacuum we  
find that $m_0 \geq 595~\rm{GeV}$.

\smallskip
\epsfysize= 2.5in
\centerline{\epsfbox{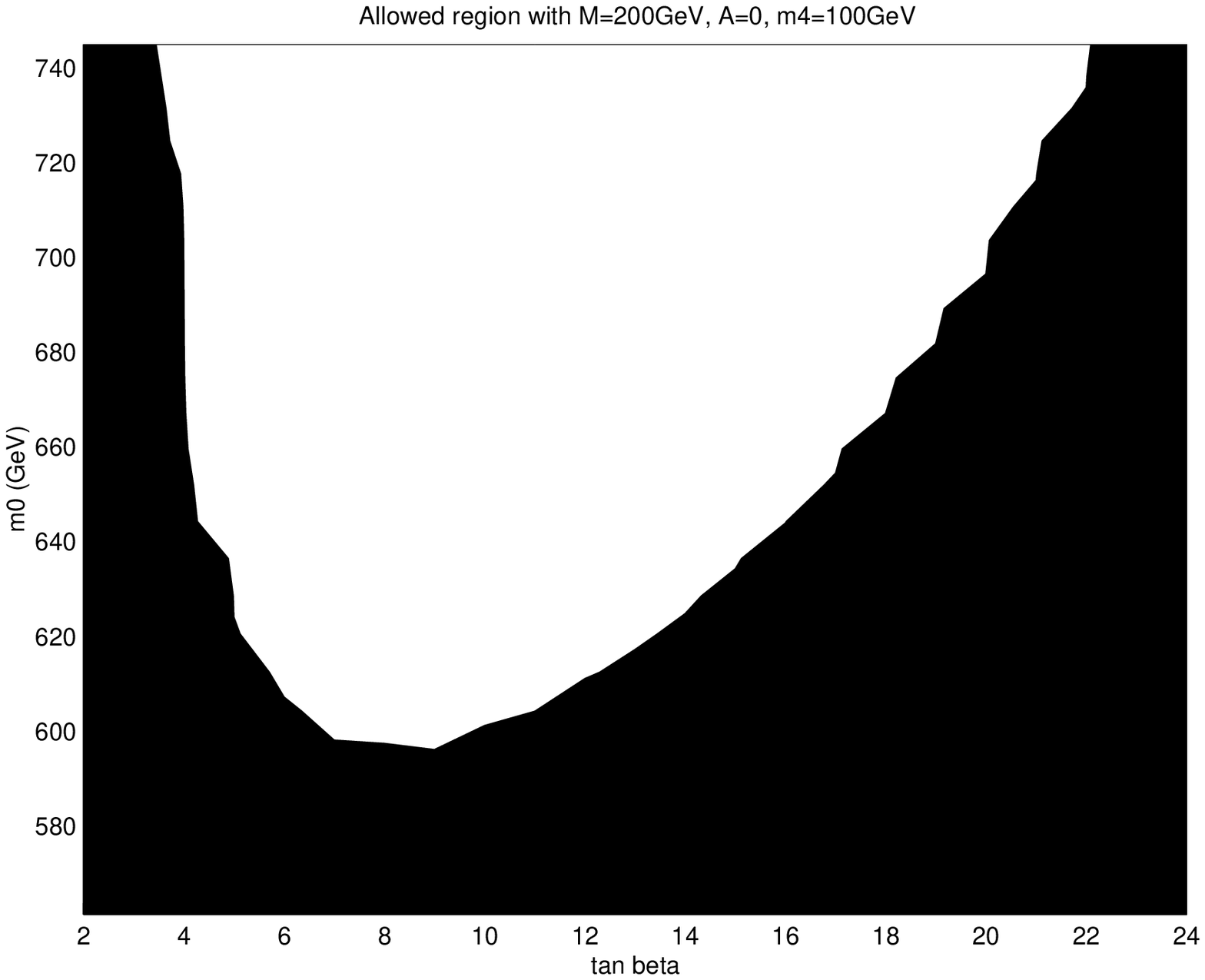}}
\in
{\it \noindent Fig.1:
The region of the $m_0, \tan\beta$ plane corresponding to an acceptable 
electroweak vacuum, for $M=200\GeV$, $m_4(M_U) = 100\GeV$ and $A = 0$. 
The shaded 
region corresponds to one or more sparticle or Higgs 
masses in violation of current 
experimental bounds.}
\medskip
\out

In Fig.~1, we show  the region of the $m_0, \tan\beta$ plane  where we
are able to obtain (by varying $m_r$)  an acceptable electroweak vacuum
for  illustrative values of $M, m_4$, and $A$. We have  made allowance
for radiative corrections by using the tree Higgs minimisation 
conditions,  but evaluated at the scale $m_0$. While a crude
approximation,  this suffices to demonstrate our main point that even
with $\mu=0$ there are  substantial regions of parameter space
available, including  ones with $m_4 < M$ and hence a Higgsino-like
light neutralino.  The lowest value of $m_0$ ($m_0\approx 590\GeV$)
corresponds to  a value of $\tan\beta\approx 8$; at this value of
$\tan\beta$ we find  that $m_{5,7,9}$ behave as in section~(4.1), i.e. 
$m_5 \approx m_9$ at $M_Z$. For $m_0 = 600\GeV$ and $\tan\beta = 8$, 
for example, 
we find $m_r = 1.06\TeV$, $m_5\approx m_9\approx 590\GeV$, 
$m_7\approx 410\GeV$, 
a Higgs with mass $84\GeV$\ and a LSP neutralino with mass $50\GeV$.   
For convenience 
we collect the sparticle mass matrices which are affected by the 
new soft breakings in an appendix.

In conclusion: if we wish to make {\it no assumptions\/} concerning the 
nature of the underlying theory, supersymmetric $\mu_{ij}$-terms should
be  replaced by the set $(m^2)^i{}_j, r^{jk}_i, m^{ij}_F, m^{ia}_A$ in
general.  With minimal unification assumptions this replaces the MSSM
$\mu$-parameter with two parameters $m_4, m_r$.  

Note added: when we submitted this paper we were unaware of 
Ref.~\ref\bfpt{F. Borzumati, G.R. Farrar, N. Polonsky and S. Thomas,
hep-ph/9902443}, in which $\phi^2\phi^*$-type soft-breakings are  used
to generate flavour mass hierarchies via radiative corrections;
and the need to consider such terms in a model--independent 
analysis was also stressed in 
Ref.~\ref\lrlh{L.J. Hall and  L. Randall, 
\prl 65 (1990) 2939}.  
We thank Nir Polonski for bringing these papers to our attention.  

\appendix{A}{The sparticle mass matrices}

In this appendix we collect the sparticle mass matrices 
which 
are affected by our generalised soft breaking. 

The stop matrix is:
\eqn\apdxa{\pmatrix{
m_Q^2 + m_t^2 + {1\over6}(4M_W^2 - M_Z^2)\cos 2\beta &
m_t ( m_{10} - m_9 \cot\beta )\cr
m_t ( m_{10} - m_9 \cot\beta ) &
m_{\tbar}^2 + m_t^2 - {2\over3}(M_W^2 - M_Z^2)\cos 2\beta\cr}.
}
Similarly for the bottom squarks we have:
\eqn\apdxb{\pmatrix{
m_Q^2 + m_b^2 - {1\over6}(2M_W^2 + M_Z^2)\cos 2\beta &
m_b ( m_8 - m_7 \tan\beta )\cr
m_b ( m_8 - m_7 \tan\beta ) &
m_{\bbar}^2 + m_b^2 + {1\over3}(M_W^2 - M_Z^2)\cos 2\beta\cr}
}
and for the tau sleptons:
\eqn\apdxc{\pmatrix{
m_L^2 + m_{\tau}^2 - {1\over2}(2M_W^2 - M_Z^2)\cos 2\beta &
m_{\tau} ( m_6 - m_5 \tan\beta )\cr
m_{\tau} ( m_6 - m_5 \tan\beta ) &
m_{\taub}^2 + m_{\tau}^2 + (M_W^2 - M_Z^2)\cos 2\beta\cr}.
}
The neutralino mass matrix is:
\eqn\apdxe{
\pmatrix{
M_1 & 0 & -M_Z\cos\beta\sin\theta_W & M_Z\sin\beta\sin\theta_W\cr
0 & M_2 & M_Z\cos\beta\cos\theta_W & - M_Z\sin\beta\cos\theta_W\cr
-M_Z\cos\beta\sin\theta_W & M_Z\cos\beta\cos\theta_W & 0 & -m_4\cr
M_Z\sin\beta\sin\theta_W & - M_Z\sin\beta\cos\theta_W & -m_4 & 0\cr
}
}  
while the chargino mass matrix is:

\eqn\apdxf{
\pmatrix{
M_2 & \sqrt{2}M_W\sin\beta\cr\sqrt{2}M_W\cos\beta & m_4\cr
}
}
The Higgs $(\hbox{mass})^2$ matrices and the sneutrino masses 
are unaffected, except inasmuch as our preferred scenario 
involves $\mu = 0$.

%
%
\listrefs   
\end